\documentclass[conference]{IEEEtran}
\IEEEoverridecommandlockouts
\usepackage{cite}
\usepackage{amsmath,amssymb,amsfonts}
\usepackage{algorithmic}
\usepackage{graphicx}
\usepackage{textcomp}
\usepackage{xcolor}
\newcommand{\chg}[1]{\textcolor{black}{#1}}

\usepackage{adjustbox}

\usepackage{multirow}
\def\BibTeX{{\rm B\kern-.05em{\sc i\kern-.025em b}\kern-.08em
    T\kern-.1667em\lower.7ex\hbox{E}\kern-.125emX}}
\begin{document}

\title{Evaluating LLM Robustness Under Domain-Specific Prompt Perturbations in Public Health Applications\\
}

\author{\IEEEauthorblockN{Chuqing Zhao}
\IEEEauthorblockA{\textit{School of Engineering and Applied Sciences} \\
\textit{Harvard University}\\
Boston, MA \\
}
\and
\IEEEauthorblockN{ Haochen Yang}
\IEEEauthorblockA{\textit{School of Engineering and Applied Sciences} \\
\textit{Harvard University}\\
Boston, MA \\
}
}

\maketitle

\begin{abstract}

Large language models (LLMs) are increasingly applied in public health applications, yet their robustness to non-clinical user inputs remains underexplored. We propose a domain-specific robustness benchmark that evaluates LLMs under two perturbation types that commonly arise when non-clinical users interact with health AI systems: misinformation framing (MF),  where prompt might be injected by false health claims, and layperson rewriting (LR), where patients describe symptoms in everyday language rather than medical terminology. Our goal is to evaluate the stability of LLMs under these perturbation. Experiments show that MF degrades accuracy by $-$7.2~pp on average with prediction flip rates of 9--38\%, even when claims are explicitly labelled as unsupported; LR causes only $-$1.4~pp degradation. These findings highlight two distinct deployment risks in public health settings: models may produce 
incorrect outputs when users unintentionally carry misinformation into their queries, and may misinterpret clinically relevant details when patients use informal language. Both risks call for perturbation-aware robustness evaluation beyond clean baseline benchmark.
\end{abstract}

\begin{IEEEkeywords}
 Natural Language Processing , Large Language Model, Robustness
\end{IEEEkeywords}

\section{Introduction}
Large language models (LLMs) are increasingly deployed in public health and clinical decision support, from answering biomedical literature questions to assisting with exam-style clinical reasoning and monitoring health-related discourse on social media ~\cite{jin2019pubmedqa, poddar2022vaccine}. Benchmarks such as PubMedQA and MedQA report strong performance on clean, professionally phrased prompts, and recent systems are often evaluated as if users always present well-formed, evidence-aligned queries. However, in real-world settings patients and the general public frequently use colloquial or imprecise health vocabulary in prompts. These conditions are not captured by standard benchmarks, leaving a critical gap in our understanding of LLM robustness under real-world public health application scenarios.

Prompt robustness, the ability of LLMs to maintain accurate and consistent output under input variation, remains underexplored in the public health domain. Existing robustness 
benchmarks such as PromptRobust~\cite{wang2023promptrobust} evaluate general-purpose NLP tasks under character-level noise  and synonym substitution, while biomedical evaluations such as 
RAmBLA~\cite{bean2024rambla} focus primarily on paraphrase variation. A few studies have examined LLM robustness in prompt injection, such as embedded misinformation, and lexical-level perturbations, such as layperson terminology, within a unified public health benchmark.

In this paper, we address this gap by evaluating four lightweighted LLMs (Llama-3.1-8B, Mistral-7B, Qwen2.5-7B, and GPT-4.1-Nano) across three public health datasets under two perturbation types: (1) misinformation framing (MF), in which a contradictory or false claim is 
injected into the prompt;  (2) layperson rewriting (LR), in which medical terminology is replaced with colloquial equivalents drawn from the Consumer Health Vocabulary~\cite{zeng2011chv}. We 
measure accuracy drop and output consistency to characterize robustness profiles across model families.
Our contributions are as follows:

\begin{itemize}
    \item \textbf{Domain-Specific Benchmark.} We construct a domain-specific robustness benchmark covering three public health tasks and two perturbation types grounded in realistic deployment scenarios.
    \item \textbf{Perturbation Comparison.} We provide a new, novel empirical comparison of misinformation framing and layperson rewriting as structurally distinct perturbation types, revealing  divergent failure patterns across multiple models.
    \item \textbf{Practical Guidance.} We provide actionable insights for practitioners by comparing open-source 7--8B models against a commercial lightweight model, offering 
    model selection guidance for resource-constrained public health settings.
\end{itemize}

\section{Related Work}
\subsection{LLMs in Public Health and Clinical NLP}
LLMs have been widely applied across public health and clinical NLP (Natural Language Processing) tasks, including biomedical question answering~\cite{jin2019pubmedqa, jin2021medqa}, vaccine stance detection~\cite{poddar2022vaccine}, and clinical decision support~\cite{thirunavukarasu2023llms,liao2025catp}. These works mainly evaluate whether LLMs can match or exceed traditional machine learning models, human annotators, or benchmark baselines under clean and well-defined input conditions. For example, Espinosa and Salath'e~\cite{espinosa2024discourse} show that GPT and open-source models, including Llama and Mistral, can extract vaccination stances from social media posts at scale. Deiner et al.~\cite{deiner2025conjunctivitis} show that LLMs such as GPT-4o and Claude can classify epidemiological characteristics from real-world outbreak-related social media posts with performance comparable to human raters. PubMedQA~\cite{jin2019pubmedqa} and MedQA-USMLE~\cite{jin2021medqa} have also become standard benchmarks for evaluating biomedical and clinical question answering.

\subsection{Prompt Robustness in Explainable NLP}
Model's robustness to prompt variation has been studies across NLP and XAI domain~\cite{sun2026beyond, han2026driftguard,han2026interpretable}. PromptRobust~\cite{wang2023promptrobust} introduces a benchmark to assess the stability of NLP models, including character-level noise, word-level substitutions, and semantic perturbations in general purpose. In public health domain, RAmBLA~\cite{bean2024rambla} shows similar sensitivity in biomedical QA, where paraphrase alone reduces PubMedQA accuracy despite preserving meaning. 

Existing robustness evaluations often assume a clinical or expert user, with perturbations typically designed by researchers with domain knowledge and inputs remaining factually accurate. This assumption overlooks important real-world scenarios in which non-clinical users may unintentionally introduce false information or describe health concerns using lay terminology. Unlike prior work that primarily examines fact-preserving perturbations or expert-authored clinical inputs, our study evaluates LLM robustness under two realistic public health input conditions: misinformation framing and layperson rewriting.

\section{Methodology}
\subsection{Problem Formulation}
Let $f(\cdot)$ denote an LLM that predicts a label $\hat{y} = f(p)$ given a prompt $p$. We define a perturbation function $\delta(\cdot)$ that modifies $p$ to produce a perturbed prompt $p' = \delta(p)$ while preserving the underlying task and ground-truth label $y$. Robustness is measured as the expected accuracy drop under perturbation:
\begin{equation}
    \Delta\text{Acc}(f, \delta) = 
    \text{Acc}(f, \mathcal{D}) - 
    \text{Acc}(f, \delta(\mathcal{D}))
\end{equation}
where $\mathcal{D}$ is the evaluation dataset. A larger 
$\Delta\text{Acc}$ indicates greater sensitivity to the 
perturbation. We additionally define the \textit{flip rate}:
\begin{equation}
    \text{Flip}(f, \delta) = 
    \frac{1}{|\mathcal{D}|}
    \sum_{p \in \mathcal{D}} 
    \mathbf{1}[f(p) \neq f(\delta(p))]
\end{equation}
which measures output instability independent of correctness.

\subsection{Perturbation Taxonomy}
We evaluate two realistic prompt perturbations that reflect noisy public-facing health communication: 

\subsubsection{\textbf{Misinformation Framing (MF)}} MF injects a false health 
claim $s_{\text{mis}}$ into prompt:
\begin{equation}
    p_{\text{MF}} = s_{\text{mis}} \oplus p_{\text{base}}
\end{equation}
where $\oplus$ denotes text concatenation. Misinformation claims are drawn from COVIDLIES~\cite{covidlies}, HealthVer~\cite{healthver}, and WHO mythbusters, retaining only statements labelled as \textit{not supported by evidence}. Each claim is matched to a sample using TF-IDF retrieval and stored in a frozen mapping file for reproducibility. We also append an explicit disclaimer to the injected claim, allowing us to test whether models can discount unsupported contextual information when warned.

\subsubsection{\textbf{Layperson Rewriting (LR)}} LR applies a term-level substitution function $\phi(\cdot)$ to the original prompt:
\begin{equation}
p_{\text{LR}} = \phi(p_{\text{base}}),
\end{equation}
where $\phi$ replaces selected medical or technical terms with colloquial equivalents drawn from the OAC Consumer Health Vocabulary~\cite{zeng2011chv}. For datasets that include an accompanying passage or context, only the question is rewritten while the passage is kept unchanged. This design isolates the effect of lexical variation in user-facing input rather than changes in the underlying evidence.

\subsection{Evaluation Metrics}
We report three evaluation metrics. \textbf{Accuracy} measures exact match between the predicted label and the ground-truth label. \textbf{$\Delta$Acc} quantifies the change in accuracy between a perturbation condition and the corresponding baseline in percentage points~(pp). \textbf{Flip rate} measures prediction instability by computing the proportion of paired examples whose predicted label changes after perturbation, regardless of whether the perturbed prediction is correct.

\section{Experimental Setup}
\subsection{Datasets}

We evaluate on three public health datasets covering distinct task types. \textbf{PubMedQA}~\cite{jin2019pubmedqa} requires answering biomedical research questions (\textit{yes / no / maybe}). \textbf{MedQA-USMLE}~\cite{jin2021medqa} consists of four-option clinical reasoning questions drawn from USMLE-style materials. \textbf{COVID-19 Vaccine Stance}~\cite{poddar2022vaccine} requires classifying tweet  toward COVID-19 vaccines as \textit{favor}, \textit{against}, or \textit{neutral}. We randomly sample 100 examples from each dataset, resulting in 3,600 inference records.

\subsection{Models}

We evaluate four lightweight LLMs spanning open-source local deployment and commercial API-based deployment. All models are evaluated with temperature set to 0 to ensure deterministic decoding. The open-source models include \textbf{Llama~3.1-8B}~\cite{llama3}, \textbf{Mistral-7B}~\cite{mistral7b}, and \textbf{Qwen2.5-7B}~\cite{qwen25}. These models are broadly comparable in scale and architecture: all are instruction-tuned, decoder-only Transformer models in the 7--8B parameter range. They are deployed locally via Ollama to support reproducible evaluation without external API dependencies. We additionally evaluate \textbf{GPT-4.1-nano}as a lightweight commercial reference model accessed through the OpenAI API ~\cite{openai2025gpt41}.

\section{Results}
\subsection{Overall Accuracy}
Table~\ref{tab:overall} reports accuracy across all models and perturbation types, averaged over three datasets. MF causes substantial accuracy degradation across all models, with open-source models dropping $\Delta$Acc $= -6.8$~pp and GPT-4.1-Nano $\Delta$Acc $= -8.0$~pp. In constrast, LR has a smaller effect: open-source models decline by only 1.8~pp on average, and GPT-4.1-Nano shows no degradation under LR ($\Delta$Acc $= 0.0$~pp).  \chg{Confidence intervals confirm these patterns: MF shifted accuracy intervals downward with non-overlapping Baseline and MF intervals, whereas LR intervals largely overlap Baseline across all models}.  This indicates that these lightweight LLMs have sufficient lexical generalization to bridge professional and lay medical terminology, but remain vulnerable to semantic-level false content injection.


\begin{table}[h]
\centering
\small
\caption{Overall accuracy (\%) \chg{with 95\% confidence intervals} and $\Delta$Acc (pp) averaged across three datasets.}
\label{tab:overall}
\begin{adjustbox}{max width=\columnwidth}
\begin{tabular}{lccccc}
\hline
\textbf{Model} & \textbf{Baseline} & \textbf{MF} & 
\textbf{LR} & \textbf{MF $\Delta$Acc} & 
\textbf{LR $\Delta$Acc} \\
\hline
Llama 3.1-8B  & \chg{69.7 [64.2--74.6]} & \chg{67.3 [61.8--72.4]} & \chg{68.0 [62.5--73.0]} & $-$2.4  & $-$1.7 \\
Mistral-7B    & \chg{57.7 [52.0--63.1]} & \chg{45.3 [39.8--51.0]} & \chg{55.0 [49.3--60.5]} & $-$12.4 & $-$2.7 \\
Qwen2.5-7B    & \chg{60.0 [54.4--65.4]} & \chg{54.0 [48.3--59.6]} & \chg{58.7 [53.0--64.1]} & $-$6.0  & $-$1.3 \\
GPT-4.1-Nano  & \chg{66.0 [60.5--71.1]} & \chg{58.0 [52.3--63.4]} & \chg{66.0 [60.5--71.1]} & $-$8.0  & $0.0$  \\
\hline
All Avg       & 63.4 & 56.2 & 61.9 & $-$7.2  & $-$1.4 \\
\hline
\end{tabular}
\end{adjustbox}
\end{table}

Each dataset represents a distinct tasks for public health domain, Table~\ref{tab:dataset} presents a per-dataset breakdown revealing patterns.

\textbf{LR has relatively smaller impact than MF.}
LR produces smaller changes than MF across all three datasets, indicating that the evaluated models are generally more robust to lexical variation than to misleading contextual information.

\textbf{MedQA is less affected by MF for most models.}
MedQA shows moderate MF degradation for Llama~3.1-8B ($-1.0$~pp), GPT-4.1-nano ($-4.0$~pp), and Qwen2.5-7B ($+2.0$~pp). One likely explanation is that the multiple-choice format constrains the output space, making it harder for unrelated misinformation to fully redirect the prediction.

\textbf{Robustness is model- and task-specific.}
The same perturbation can affect different models very differently. Llama~3.1-8B remains stable across tasks, while Mistral-7B shows large MF drops on both PubMedQA and MedQA. Qwen2.5-7B is relatively stable on MedQA but highly vulnerable on Vaccine Stance. These patterns indicate that clean benchmark accuracy alone is insufficient for deployment decisions. Robustness should be evaluated separately by task and perturbation type.

\begin{table}[h]
\centering
\small
\caption{Accuracy (\%) and $\Delta$Acc (pp) by model, 
condition, and dataset.}
\label{tab:dataset}
\setlength{\tabcolsep}{4pt}
\begin{tabular}{llcccc}
\hline
\textbf{Model} & \textbf{} & 
\textbf{PubMed} & \textbf{MedQA} & 
\textbf{Vaccine} & \textbf{Avg.} \\
\hline
\multirow{5}{*}{Llama 3.1-8B}
 & Baseline       & 76.0 & 60.0 & 73.0 & 69.7 \\
 & MF         & 73.0 & 59.0 & 70.0 & 67.3 \\
 & LR         & 76.0 & 56.0 & 72.0 & 68.0 \\
 & $\Delta$MF & $-$3.0 & $-$1.0 & $-$3.0 & $-$2.4 \\
 & $\Delta$LR & $0.0$ & $-$4.0 & $-$1.0 & $-$1.7 \\
\hline
\multirow{5}{*}{Mistral-7B}
 & Baseline       & 54.0 & 50.0 & 69.0 & 57.7 \\
 & MF         & 44.0 & 39.0 & 53.0 & 45.3 \\
 & LR         & 52.0 & 46.0 & 67.0 & 55.0 \\
 & $\Delta$MF & $-$10.0 & $-$11.0 & $-$16.0 & $-$12.4 \\
 & $\Delta$LR & $-$2.0 & $-$4.0 & $-$2.0 & $-$2.7 \\
\hline
\multirow{5}{*}{Qwen2.5-7B}
 & Baseline       & 44.0 & 63.0 & 73.0 & 60.0 \\
 & MF         & 43.0 & 65.0 & 54.0 & 54.0 \\
 & LR         & 42.0 & 61.0 & 73.0 & 58.7 \\
 & $\Delta$MF & $-$1.0 & $+$2.0 & $-$19.0 & $-$6.0 \\
 & $\Delta$LR & $-$2.0 & $-$2.0 & $0.0$ & $-$1.3 \\
\hline
\multirow{5}{*}{GPT-4.1-Nano}
 & Baseline       & 52.0 & 72.0 & 74.0 & 66.0 \\
 & MF         & 43.0 & 68.0 & 63.0 & 58.0 \\
 & LR         & 50.0 & 73.0 & 75.0 & 66.0 \\
 & $\Delta$MF & $-$9.0 & $-$4.0 & $-$11.0 & $-$8.0 \\
 & $\Delta$LR & $-$2.0 & $+$1.0 & $+$1.0 & $0.0$ \\
\hline
\end{tabular}
\end{table}

\subsection{Flip Rate Analysis}
\begin{figure}[h]
    \centering
    \includegraphics[width=1\columnwidth]
    {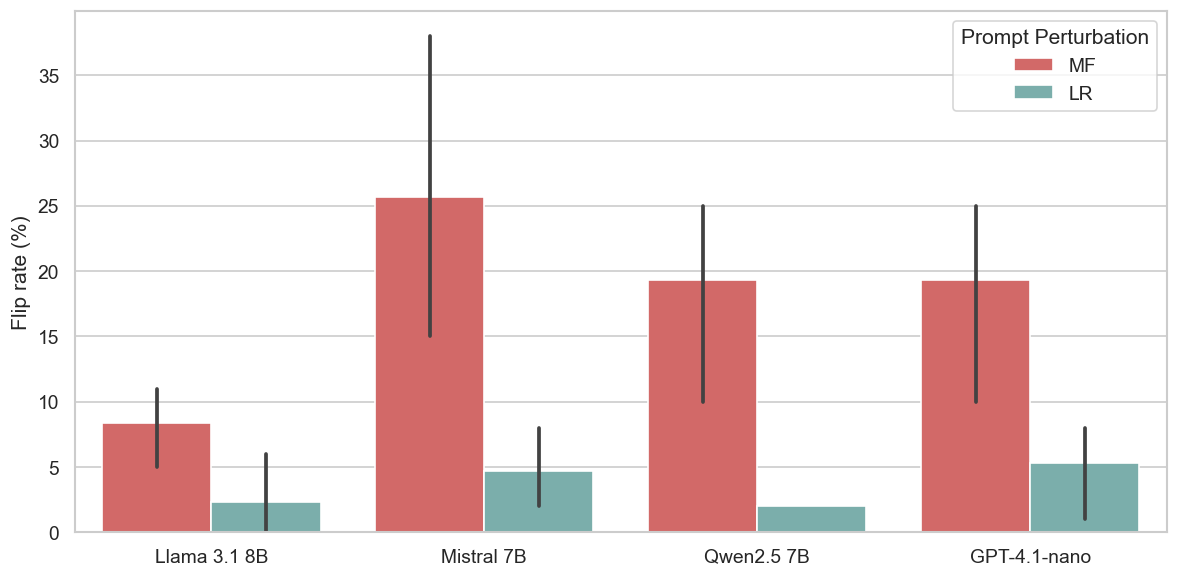}
    \caption{Flip rate (\%) under MF and LR 
    perturbations, averaged across three datasets. 
    Error bars indicate variance across datasets.}
    \label{fig:flip}
\end{figure}

Figure~\ref{fig:flip} reports the average flip rate across three datasets: the proportion of samples on which the predicted label changes between baseline and perturbed conditions.

\textbf{ Flip rate and accuracy drop are decoupled.}
Qwen2.5-7B and GPT-4.1-Nano share similar MF flip rates ($\sim$19.5\%) yet differ substantially in accuracy drop ($-$6.0 vs.\ $-$8.0~pp), indicating that GPT-4.1-Nano flips more correct predictions whereas Qwen2.5-7B more frequently switches between incorrect labels.

\textbf{Robustness variability reveals task-specific risk.} Mistral-7B has the widest MF flip-rate range across tasks (15--38\%), while Llama~3.1-8B has the narrowest range (5--11\%). This indicates that Mistral-7B's robustness is highly task-dependent: it may be acceptable on some tasks but becomes unstable on other.

\subsection{Qualitative Analysis}
To complement the quantitative results, we examine representative cases illustrating the primary 
failure modes observed under MF and LR perturbations.
Table~\ref{tab:cases} summarizes each case; detailed 
discussion follows.

\begin{table*}[t]
\centering
\small
\caption{Representative qualitative cases. 
$\checkmark$ = correct; $\times$ = incorrect.}
\label{tab:cases}
\setlength{\tabcolsep}{5pt}
\begin{tabular}{clp{3.9cm}p{3.5cm}cccc}
\hline
\textbf{\#} & \textbf{Dataset} & \textbf{Input (truncated)} & 
\textbf{Injected Claim (truncated)} & 
\textbf{Llama} & \textbf{Mistral} & 
\textbf{Qwen} & \textbf{GPT} \\
\hline
1 & Vaccine   
  & \textit{If this holds up, it would be so much better than expected... \#vaccine} (GT: favor)
  & Natural immunity is always better than vaccine-induced immunity
  & B$\to$N\,$\times$ & B$\to$N\,$\times$ & B$\to$A\,$\times$ & B$\to$N\,$\times$ \\
\hline
2 & PubMedQA  
  & \textit{Gender difference in survival of resected non-small cell lung cancer?} (GT: yes)
  & ICU triage claim
  & \textbf{Y$\to$Y\,$\checkmark$} & Y$\to$M\,$\times$ & Y$\to$M\,$\times$ & Y$\to$M\,$\times$ \\
\hline
3 & MedQA    
  & SLE patient with DVT (GT: B); LR replaces \textit{Prothrombin Time}, \textit{Platelets}{\ldots}
  & --- (LR only)
  & B$\to$A\,$\times$ & B$\to$A\,$\times$ & \textbf{B$\to$B\,$\checkmark$} & \textbf{B$\to$B\,$\checkmark$} \\
\hline
\end{tabular}
\begin{flushleft}
\footnotesize
B = baseline prediction; N = neutral; M = maybe; 
Y = yes; A/D = MCQ options. 
Bold = model maintains or recovers correct prediction under perturbation.
\end{flushleft}
\end{table*}

\noindent\textbf{Cases 1: MF overrides correct 
predictions}
All four models answer correctly at baseline but flip under MF. The injected claim is directly relevant to vaccine efficacy, making it highly salient for stance classification and causing models to misclassify a favorable tweet.

\noindent\textbf{Case 2: Hedging as a failure mode on PubMedQA.}
Three models shift from \textit{yes} to \textit{maybe} under a weakly related MF claim. This suggests that MF does not always push models toward the opposite answer; it can instead increase uncertainty, which still lowers accuracy when the ground truth is \textit{yes}.

\noindent\textbf{Case 3: LR breaks clinical 
reasoning chains on MedQA.}
Although LR has small average impact, this case shows that rewriting diagnostic test names and answer terminology in LR can disrupt cross-referencing between the question and options.

\section{Discussion}
Three key findings emerge from our evaluation.
\textbf{MF and LR pose two distinct challenges.} MF poses a substantially greater robustness threat than LR ($-7.2$~pp vs.\ $-1.4$~pp average), with flip rates of
9--38\%. MF degradation of -\$7.2 pp implies approximately one additional error per 14 queries, a non-trivial risk in public health decision support. . LR shows a smaller decline ($-$1.4 pp, or $\sim$1 error per 71 queries). This suggests models over-weight contextual content over 
parametric knowledge, consistent with findings on LLM 
sycophancy~\cite{chen2025sycophancy}.

\textbf{Task structure shapes vulnerability.}
Vaccine Stance, which lacks an evidence passage or structured
clinical answer options to anchor predictions, suffers the
largest MF degradation ($-3.0$ to $-19.0$~pp). In contrast,
PubMedQA provides an evidence passage and MedQA constrains
the output space through multiple-choice options, which may
partially reduce the influence of injected misinformation.

\textbf{Baseline accuracy does not guarantee robustness.}
High baseline performance does not necessarily imply
perturbation resistance: GPT-4.1-nano achieves 66.0\%
baseline accuracy but drops $-8.0$~pp under MF, over three
times the degradation of Llama~3.1-8B ($-2.4$~pp).

\textbf{Limitations.}
Each dataset uses 100 samples; MF uses a single severity
level; and LR coverage is bounded by the CHV vocabulary.
Future work will introduce severity-graded MF, additional
model families, and defense mechanisms.

\bibliographystyle{IEEEtran}
\bibliography{references}

@inproceedings{jin2019pubmedqa,
  author    = {Jin, Qiao and Dhingra, Bhuwan and Liu, Zhiyong 
               and Cohen, William W. and Lu, Xinghua},
  title     = {{PubMedQA}: A Dataset for Biomedical Research 
               Question Answering},
  booktitle = {Proceedings of the 2019 Conference on Empirical 
               Methods in Natural Language Processing (EMNLP)},
  year      = {2019},
  pages     = {2567--2577},
  publisher = {Association for Computational Linguistics},
  doi       = {10.18653/v1/D19-1259}
}

@inproceedings{jin2021medqa,
  author    = {Jin, Di and Pan, Eileen and Oufattole, Nassim 
               and Weng, Wei-Hung and Fang, Hanyi and Szolovits, Peter},
  title     = {What Disease Does This Patient Have? A Large-Scale 
               Open Domain Question Answering Dataset from Medical Exams},
  booktitle = {Applied Sciences},
  volume    = {11},
  number    = {14},
  pages     = {6421},
  year      = {2021},
  publisher = {MDPI},
  doi       = {10.3390/app11146421}
}

@inproceedings{poddar2022vaccine,
  author={Poddar, Soham and Mondal, Mainack and Misra, Janardan and Ganguly, Niloy and Ghosh,               Saptarshi},
  title     = {Winds of Change: Impact of {COVID-19} on Vaccine-Related 
               Opinions of Twitter Users},
  booktitle = {Proceedings of the international aaai conference on web and social media},
  volume={16},
  pages={782--793},
  year      = {2022},
  doi       = {10.1609/icwsm.v16i1.19334}
}

@inproceedings{wang2023promptrobust,
    author = {Zhu, Kaijie and Wang, Jindong and Zhou, Jiaheng and Wang, Zichen and Chen, Hao and Wang, Yidong and Yang, Linyi and Ye, Wei and Zhang, Yue and Gong, Neil and Xie, Xing},
    title = {PromptRobust: Towards Evaluating the Robustness of Large Language Models on Adversarial Prompts},
    year = {2024},
    publisher = {ACM},
    doi = {10.1145/3689217.3690621},
    booktitle = {Proceedings of the 1st ACM Workshop on Large AI Systems and Models with Privacy and Safety Analysis},
    pages = {57–68},
    }

@article{bean2024rambla,
  author    = {Bolton, William James and Poyiadzi, Rafael and Morrell, Edward R and Bueno,                      Gabriela van Bergen Gonzalez and Goetz, Lea},
  title     = {RAmBLA: A framework for evaluating the reliability of LLMs as assistants in the                  biomedical domain},
  journal   = {arXiv preprint arXiv:2403.14578},
  year      = {2024},
  doi       = {10.48550/arXiv.2403.14578}
}

@article{zeng2011chv,
  author    = {Zeng-Treitler, Qing and Goryachev, Sergey and 
               Kim, Hyeoneui and Keselman, Alla and Rosendale, Deborah},
  title     = {Making Texts in Electronic Health Records Comprehensible 
               to Consumers: A Prototype Translator},
  journal   = {AMIA Annual Symposium Proceedings},
  volume    = {2007},
  pages     = {836--840},
  year      = {2007}
}

@article{espinosa2024discourse,
  author  = {Espinosa, Laura and Salath{\'e}, Marcel},
  title   = {Use of Large Language Models as a Scalable Approach 
             to Understanding Public Health Discourse},
  journal = {PLOS Digital Health},
  volume  = {3},
  number  = {10},
  pages   = {e0000631},
  year    = {2024},
  doi     = {10.1371/journal.pdig.0000631}
}

@article{deiner2025conjunctivitis,
  author  = {Deiner, Michael S. and others},
  title   = {Use of Large Language Models to Classify 
             Epidemiological Characteristics in Synthetic and 
             Real-World Social Media Posts About Conjunctivitis 
             Outbreaks: Infodemiology Study},
  journal = {Journal of Medical Internet Research},
  volume  = {27},
  pages   = {e65226},
  year    = {2025},
  doi     = {10.2196/65226}
}

@article{llama3,
  author  = {Dubey, Abhimanyu and others},
  title   = {The {Llama} 3 Herd of Models},
  journal = {arXiv preprint arXiv:2407.21783},
  year    = {2024},
  doi     = {10.48550/arXiv.2407.21783}
}

@article{mistral7b,
  author  = {Jiang, Albert Q. and others},
  title   = {Mistral {7B}},
  journal = {arXiv preprint arXiv:2310.06825},
  year    = {2023},
  doi     = {10.48550/arXiv.2310.06825}
}

@article{qwen25,
  author  = {{Qwen} and Yang, An and Yang, Baosong and Zhang, Beichen 
             and Hui, Binyuan and others},
  title   = {{Qwen2.5} Technical Report},
  journal = {arXiv preprint arXiv:2412.15115},
  year    = {2024},
  doi     = {10.48550/arXiv.2412.15115}
}

@inproceedings{covidlies,
    title = "{COVIDL}ies: Detecting {COVID}-19 Misinformation on Social Media",
    author = "Hossain, Tamanna  and
      Logan IV, Robert L.  and
      Ugarte, Arjuna  and
      Matsubara, Yoshitomo  and
      Young, Sean  and
      Singh, Sameer",
    editor = "Verspoor, Karin  and
      Cohen, Kevin Bretonnel  and
      Conway, Michael  and
      de Bruijn, Berry  and
      Dredze, Mark  and
      Mihalcea, Rada  and
      Wallace, Byron",
    booktitle = "Proceedings of the 1st Workshop on NLP for COVID-19 (Part 2) at EMNLP 2020",
    month = dec,
    year = "2020",
    address = "Online",
    publisher = "Association for Computational Linguistics",
    url = "https://aclanthology.org/2020.nlpcovid19-2.11",
    doi       = {10.18653/v1/2020.nlpcovid19-2.11}
}

@inproceedings{healthver,
    title = "Evidence-based Fact-Checking of Health-related Claims",
    author = "Sarrouti, Mourad  and
      Ben Abacha, Asma  and
      Mrabet, Yassine  and
      Demner-Fushman, Dina",
    editor = "Moens, Marie-Francine  and
      Huang, Xuanjing  and
      Specia, Lucia  and
      Yih, Scott Wen-tau",
    booktitle = "Findings of the Association for Computational Linguistics: EMNLP 2021",
    month = nov,
    year = "2021",
    address = "Punta Cana, Dominican Republic",
    publisher = "Association for Computational Linguistics",
    url = "https://aclanthology.org/2021.findings-emnlp.297/",
    doi = "10.18653/v1/2021.findings-emnlp.297",
    pages = "3499--3512",
}

@article{thirunavukarasu2023llms,
    title = "Large language models in medicine",
    author = "Thirunavukarasu, Ajay J. and others",
    journal = "Nature Medicine",
    volume = "29",
    number = "8",
    pages = "1930--1940",
    year = "2023",
    doi = "10.1038/s41591-023-02448-8",
    url = "https://doi.org/10.1038/s41591-023-02448-8",
}

@misc{openai2025gpt41,
  author       = {{OpenAI}},
  title        = {Introducing {GPT-4.1} in the {API}},
  year         = {2025},
  month        = apr,
  howpublished = {\url{https://openai.com/index/gpt-4-1/}},
  note         = {Accessed: 2025}
}

@misc{chen2025sycophancy,
    title = "{SycEval}: Evaluating {LLM} Sycophancy",
    author = "Fanous, Aaron and
      Goldberg, Jacob and
      Agarwal, Ank A. and
      Lin, Joanna and
      Zhou, Anson and
      Daneshjou, Roxana and
      Koyejo, Sanmi",
    year = "2025",
    month = feb,
    note = "AIES 2025",
    doi = "10.48550/arXiv.2502.08177",
    url = "https://arxiv.org/abs/2502.08177",
}

@inproceedings{liao2025catp,
  author    = {Liao, Ruqi and Zhao, Chuqing and Li, Jin and 
               Feng, Weiqi and Lyu, Yi and Chen, Bingxian and 
               Yang, Haochen},
  title     = {{CATP}: Cross-Attention Token Pruning for 
               Accuracy Preserved Multimodal Model Inference},
  booktitle = {2025 IEEE Conference on Artificial Intelligence 
               (CAI)},
  pages     = {1100--1104},
  year      = {2025},
  publisher = {IEEE},
  doi       = {10.1109/CAI63862.2025.11050741}
}

@article{sun2026beyond,
  title={Beyond Accuracy: Measuring Bias Acknowledgment in Chain-of-Thought Reasoning for Responsible AI Evaluation},
  author={Sun, Xian and Gao, Wei and Wang, Yingshuo and Kong, Lingdong and Li, Yanhang and Fan, Zhichao and Zhuang, Zexin and Dong, Wenlong and Zheng, Zhiyuan and Paranjape, Hrishikesh and others},
  journal={arXiv preprint arXiv:2606.15127},
  year={2026}
}

@article{han2026driftguard,
  title={DriftGuard: Mitigating Asynchronous Data Drift in Federated Learning},
  author={Han, Yizhou and Wu, Di and Varghese, Blesson},
  journal={arXiv preprint arXiv:2603.18872},
  year={2026}
}

@article{han2026interpretable,
  title={Interpretable vs Learned Encoders for High-Cardinality Fraud Detection},
  author={Han, Xiao and Liu, Jingjing and Zheng, Moxuan and Zhang, Zhen and Wu, Chenyu},
  journal={arXiv preprint arXiv:2607.00477},
  year={2026}
}

\end{document}